# The use of deep learning enables high diagnostic accuracy in detecting syndesmotic instability on weight-bearing CT scanning

Alireza Borjali[1,2,*], Soheil Ashkani-Esfahani[2,3], Rohan Bhimani[3], Daniel Guss[2,3], Orhun K. Muratoglu[1,2], Christopher W. DiGiovanni[2,3], Kartik Mangudi Varadarajan[2], Bart Lubberts[3]

[1]Harris Orthopaedics Laboratory, Department of Orthopaedic Surgery, Massachusetts General Hospital, Boston, MA, USA
[2]Department of Orthopaedic Surgery, Harvard Medical School, Boston, MA, USA
[3]Foot & Ankle Research and Innovation Laboratory, Massachusetts General Hospital, Harvard Medical School, Boston, MA, USA

*Corresponding author: Alireza Borjali
aborjali@mgh.harvard.edu
a.borjali@gmail.com

**ABSTRACT**

**Background**

Delayed diagnosis of syndesmosis instability can lead to significant morbidity and accelerated arthritic change in the ankle joint. Weight-bearing computed tomography (WBCT) has shown promising potential for early and reliable detection of isolated syndesmotic instability using 3D volumetric measurements. While these measurements have been reported to be highly accurate, they are also experience-dependent, time-consuming, and need a particular 3D measurement software tool that leads the clinicians to still show more interest in the conventional diagnostic methods for syndesmotic instability.

**Purpose**

The purpose of this study was to increase accuracy, accelerate analysis time, and reduce inter-observer bias by automating 3D volume assessment of syndesmosis anatomy using WBCT scans.

**Methods**

We conducted a retrospective study using previously collected WBCT scans of patients with unilateral syndesmotic instability. 144 bilateral ankle WBCT scans were evaluated (48 unstable, 96 control). We developed three deep learning (DL) models for analyzing WBCT scans to recognize syndesmosis instability. These three models included two state-of-the-art models (Model 1 - 3D convolutional neural network [CNN], and Model 2 - CNN with long short-term memory [LSTM]), and a new model (Model 3 - differential CNN LSTM) that we introduced in this study

**Results**

Model 1 failed to analyze the WBCT scans (F1-score = 0). Model 2 only misclassified two cases (F1-score = 0.80). Model 3 outperformed Model 2 and achieved a nearly perfect performance, misclassifying only one case (F1-score = 0.91) in the control group as unstable while being faster than Model 2.

**Conclusion**

We have developed a DL model for 3D WBCT syndesmosis assessment that achieved very high accuracy and accelerated analytics. This DL model shows promise for use by clinicians to improve diagnostic accuracy, reduce measurement bias, and save both time and expenditure for the healthcare system.

## 1. INTRODUCTION

Syndesmotic instability, left un or misdiagnosed, can lead to long-term complications including pain, limited ankle range of motion, and degenerative changes of the ankle joint[1,2]. Given that syndesmosis instability is a three-dimensional phenomenon—occurring in the coronal, sagittal, and rotational plane—evaluating the joint as such may help with a more accurate diagnosis[3,4]. Bilateral weight-bearing computed tomography (WBCT) has recently received much attention for diagnosing syndesmotic instability mainly due to its ability to provide a dynamic bilateral 3D view of the ankle joint under physiologic weight[3–6]. Ashkani Esfahani et al. recently reported 3D WBCT volume measurement of the syndesmosis is to be a very reliable technique for diagnosing syndesmotic instability, with an accuracy of 90%, a sensitivity of 95.8% (95% confidence interval (CI):87.8-100), and a specificity of 83.3% (95% CI: 68.4-98.2)[5]. Despite the relatively high accuracy of this method, however, manual 3D volume measurement proved also to be fairly time-consuming, vulnerable to measurement bias, and dependent on specific software tools. Alternative use of a computer-assisted automated 3D analysis of the WBCT images that can also compare sides to detect instability, though, could potentially hasten diagnosis, reduce measurement bias, and improve overall accuracy.

Deep learning (DL), an emerging powerful branch of machine learning, has yielded numerous breakthroughs for detecting and characterizing musculoskeletal pathologies [7–12] with potential application in all other image dependent diagnosis such as skin maladies[13–15]. DL models possess the ability to learn semantically meaningful patterns in imaging data using a specific algorithm without being explicitly programmed by humans. This algorithm can iteratively adjust and readjust its neural network connections after being repeatedly exposed to input data and the desired outcomes. Training a DL model using a sufficient image database enables the algorithm to distinguish normal from abnormal images in a fraction of a second with high accuracy and reproducibility.

The objective of this research was to train three DL models using bilateral WBCT scans of patients with unilateral syndesmotic instability and compare their performance in the detection of instability. We hypothesized that DL models can detect an unstable syndesmosis with far greater speed and accuracy than more traditional human interpretation.

## 2. METHODS

*2.1 Study Design*

After acquiring institutional review board (IRB) approval, a retrospective study was conducted using previously collected WBCT data.

*2. 2 Data*

A total of 144 bilateral ankle WBCT scans were evaluated (48 unstable, 96 control). These WBCTs were acquired from patients seen between 2016-2020 at an academic hospital. Images were obtained using a PedCAT™ device (CurveBeam, Warrington, PA). The exclusion criteria were patients younger than 18 years old, not able to bear weight, those with bilateral ankle injuries, ipsilateral fracture of the distal tibia with extension to the syndesmotic space or posterior tibial tubercle. Demographic and injury data of included patients are shown in Table 1. Notably, among patients in the unstable group, WBCT scans were taken 12.6 ± 13.6 months after the initial trauma, suggesting that most injuries were chronic. The control group included participants with stable ankle and syndesmosis and who had undergone WBCT to assess the tarsometatarsal joints, sesamoids, bone fragments, metatarsophalangeal malalignments (i.e. Lisfranc instability), or evaluation of the forefoot postoperatively.

**Table 1.** Demographic characteristics of individuals in the patients and the control group.

|  | Control group (N=96) |  | Patient group (N=48) |  | p-value |
| --- | --- | --- | --- | --- | --- |
| Gender | Male | 39.6% (n=38) | Male | 45.8% (n=22) | 0.22† |
|  | Female | 60.4% (n=58) | Female | 54.2% (n=26) |  |
| Age (years; mean ± SD) | 31.5±19.1 |  | 33.8±11.9 |  | 0.63‡ |
| BMI (kg/m$^2$; mean ± SD) | 29.8±4.9 |  | 27.1±5.9 |  | 0.45‡ |

† Chi-square test, P<0.05 considered as statistically significant
‡ Student T-test, P<0.05 considered as statistically significant
**Abbreviations:** BMI, body mass index; SD, standard deviation.

*2.3 Ground Truth*

Syndesmosis instability was confirmed intraoperatively using arthroscopic assessment or via direct examination in open surgery. Following diagnosis, all patients received surgical fixation of the syndesmosis.

*2.4 Data Preparation*

All WBCT scans were acquired from the Institutional Patient Data Registry (RPDR) as Digital Imaging and Communications in Medicine (DICOM) files. Prior to analyses, the DICOM files were de-identified. On bilateral ankle WBCT images taken in the axial plane, a total of thirteen CT slices with a slice thickness of 0.3 mm starting from the tibial plafond up to 5cm proximal to the syndesmoses were selected in both the unstable and control groups (Fig. 1). We choose to select thirteen slices to keep the number of inputs to the deep learning models constant while making sure that the entire syndesmosis is captured within these slices. Subsequently, all CT slices were translated from DICOM to JPEG format. JPEG images were then rescaled to 224 by 224 pixels and were normalized by dividing each pixel by the maximum pixel value of 255. The entire data set (144 individuals, 13 images per individual) was randomly divided into the training, validation, and the final test subsets with an 80:10:10 split ratio. This split ratio was maintained among the unstable and control groups to ensure that each group had the same number of representatives in the training, validation, and test subsets.

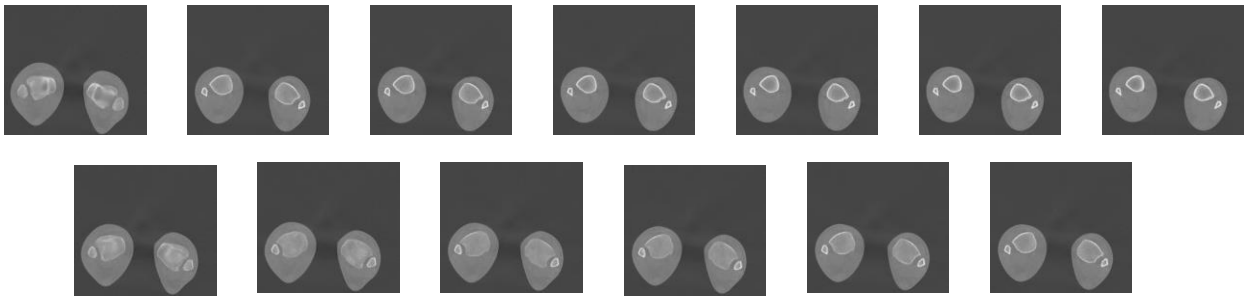

**Fig. 1** An example weight-bearing computed tomography (WBCT) scan showing the thirteen selected slices with a slice thickness of 0.3 mm starting from the tibial plafond up to 5cm proximal to the syndesmoses

*2.5 Models*

A WBCT scan consists of a stack of sequential slices. The position of each slice in the WBCT scan stack is important when analyzing the WBCT scan. We developed three DL models to analyze the WBCT images. The position of each slice in the stack and the diagnostic information that could be gained from each slice were captured by comparing one slice to the next slices and the previous slices. We developed these models using state-of-the-art methods (Model 1 and Model 2) and a new method that we propose (Model 3) for analyzing sequential medical images in this study.

*2.5.1 Model 1: 3D Convolutional Neural Network*

Convolutional neural networks (CNNs) are mainly used for analyzing 2D images with 2D convolution layers. However, 3D convolution layers can be used in a CNN structure to capture features in both the spatial and temporal dimensions. We developed Model 1 with a 3D convolution layer that convolved over the entire stack of a given WBCT scan to extract features in both spatial and temporal dimensions (Fig. 2). This model essentially "looked" at the entire WBCT stack at once to make a decision and categorize a given WBCT scan as "unstable" or "control." Details of Model 1 are shown in Fig. 2. We used random Gaussian distribution weights to initialize this model.

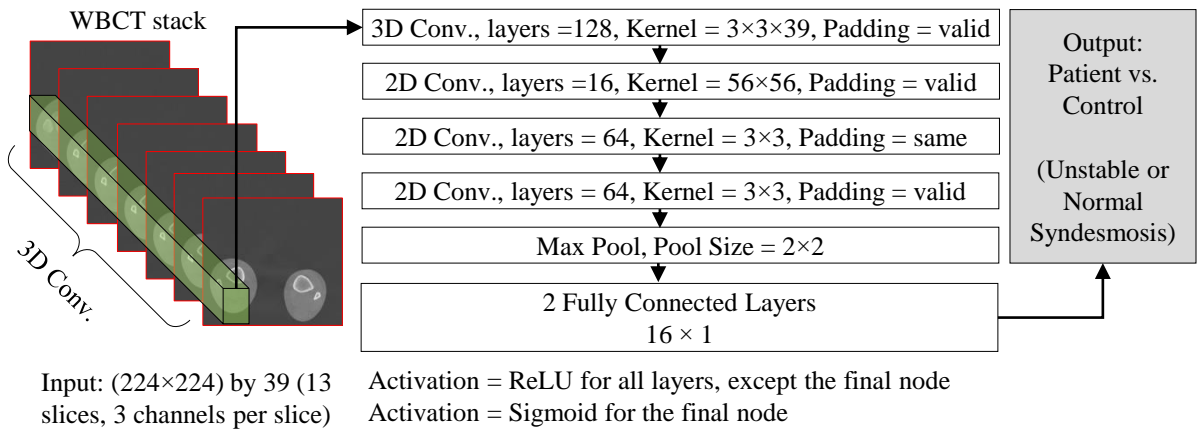

**Fig. 2** Model 1 architecture: 3D Convolutional Neural Network (CNN)

*2.5.2 Model 2: Convolutional Neural Network -Long Short-Term Memory*

Recurrent Neural Networks (RNN) are a class of neural networks designed for handling sequential data. In an RNN model, the output of each step is a function of the output of the previous steps. In other words, an RNN model has the "memory" to look back at what has been calculated so far at each step and to incorporate that into its calculation for the current step. Long Short-Term Memory (LSTM) is a variation of vanilla RNN to capture long-range dependencies in sequential data. Another variation of LSTM is a multilayer LSTM. Multilayer LSTM stacks layers of LSTMs on top of each other to create a deeper model and increase the overall computation capacity for more complicated tasks.

We combined a CNN model with a multilayer LSTM model to create Model 2 (Fig. 3). The CNN model was used to analyze each slice individually and summarize each slice's features. We used a VGG-16 CNN that was pre-trained on the ImageNet dataset as a feature extractor. Subsequently, those features were fed into the LSTM network. The LSTM network remembered the features of all the slices that came before a given slice when it was analyzing that slice. This way the spatial information of the individual slices was captured and were considered in the overall calculation.

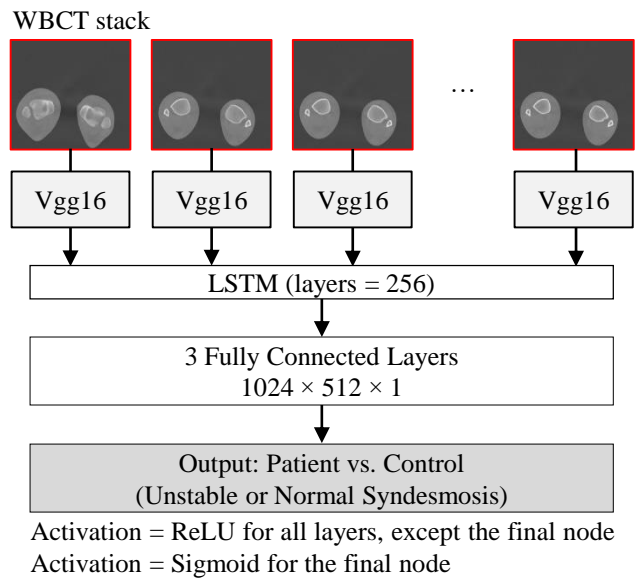

**Fig. 3** Model 2 architecture: Convolutional Neural Network -Long Short-Term Memory (CNN-LSTM)

*2.5.3 Model 3: Differential Convolutional Neural Network -Long Short-Term Memory (DCNN-LSTM)*

In this model, we used the same structure as Model 2 but used changes between successive sequential slices as input (Fig. 4). The input to this DL model was pixel-wise subtraction of successive slices ($\Delta I$) defined as follows *(eq. 1)*:

$$(eq.1)\ \Delta I = Slice_i - Slice_{i-1}$$

This model used ΔI as input as opposed to the raw image slices as was done in Model 2. By taking ΔI as input, this model could more efficiently focus on the region of interest, in particular those aspects of the images that changed between the two image slices. This approach also aided in reducing the input "noise"/ "background" information, which was not relevant for the task at hand.

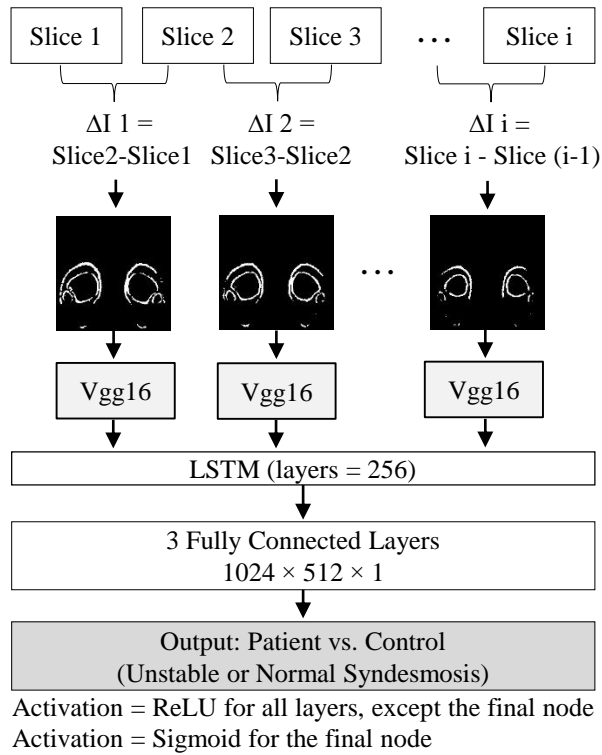

**Fig. 4** Model 3 architecture: Differential Convolutional Neural Network -Long Short-Term Memory (DCNN-LSTM)

*2.6 Training*

All models were trained using Adam optimizer (initial learning rate = 0.001, beta 1 = 0.9, beta 2 = 0.999, epsilon = 1 e-8), with a batch size of 5 for 1000 epochs with early stoppage criteria. The models were implemented with Tensorflow (version r1.9) with Keras (version 2.2.0) backend on Python (version 3.6) running on a workstation comprised of an Intel(R) Xeon(R) Gold 6128 processor, 64GB of DDR4 RAM, and an NVIDIA Quadro P5000 graphic card.

*2.7 Evaluation*

After achieving optimum performance on the validation subset the models were tested on the holdout test subset, which was isolated from the training process, and the results were reported as the models' performance. We reported a confusion matrix for each model calculating sensitivity, specificity, accuracy, and F1-score with an optimum threshold. We also plotted the receiver operating characteristic (ROC) curve and calculated the area under the curve (AUC) for all models. Moreover, we reported the processing time per epoch for each model.

**3. RESULTS**

Figure 5 shows ROC curves with AUC for all models identifying the patients with an unstable syndesmosis in the test subset. Model 1 (3D CNN) could not identify any patient (AUC = 0.0) while Model 2 (CNN-LSTM) and Model 3 (DCNN-LSTM) had high performances (AUC of 0.80 and 0.91, respectively).

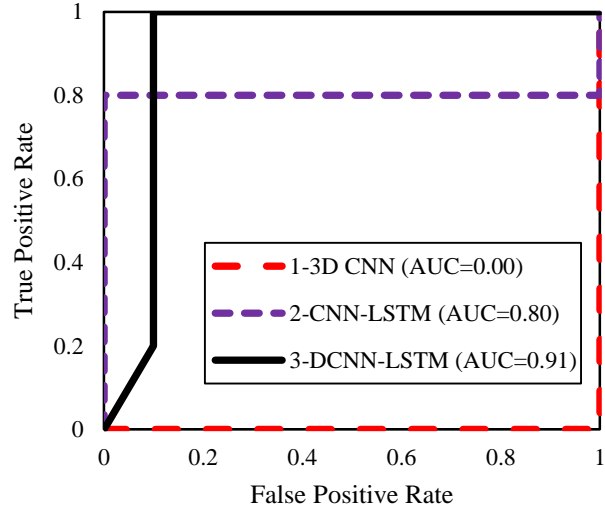

**Fig. 5** Receiver operating characteristic (ROC) curve showing performance and area under the curve (AUC) for all models including Model 1: 3D Convolutional Neural Network (3 D CNN), Model 2: Convolutional Neural Network -Long Short-Term Memory (CNN-LSTM), and Model 3: Differential Convolutional Neural Network -Long Short-Term Memory (DCNN-LSTM)

Table 1 shows the performance of all models at the optimum threshold identified from Fig. 1. Table 1 shows that Model 2 (CNN-LSTM) only misclassified two cases (F1-score = 0.80). On the other hand, Model 3 (DCNN-LSTM) outperformed Model 2 and achieved a nearly perfect performance misclassifying only one case (F1-score = 0.91) in the control group as unstable.

**Table 1.** Results of different models for classifying the weight-bearing computed tomography (WBCT) scans in the test subset including Model, 1 3D Convolutional Neural Network (3D CNN), Model 2, Convolutional Neural Network-Long Short-Term Memory (CNN-LSTM), and Model 3, Differential Convolutional Neural Network-Long Short-Term Memory (DCNN-LSTM)

| Model | | | True Unstable | True Control | | |
|---|---|---|---|---|---|---|
| | **1- 3D CNN** | | True Unstable | True Control | Sensitivity | 0.00 |
| | | Predicted Unstable | 0 | 0 | Specificity | 1.00 |
| | | Predicted Control | 5 | 10 | Accuracy | 0.67 |
| | | | | | F1-score | 0.00 |
| | **2- CNN-LSTM** | | True Unstable | True Control | Sensitivity | 0.80 |
| | | Predicted Unstable | 4 | 1 | Specificity | 0.90 |
| | | Predicted Control | 1 | 9 | Accuracy | 0.87 |
| | | | | | F1-score | 0.80 |
| | **3- DCNN-LSTM** | | True Unstable | True Control | Sensitivity | 1.00 |
| | | Predicted Unstable | 5 | 1 | Specificity | 0.90 |
| | | Predicted Control | 0 | 9 | Accuracy | 0.93 |
| | | | | | F1-score | 0.91 |

Figure 6 shows each model's average processing time per epoch. Model 1 (3D CNN) was the fastest one; however, it had the lowest performance as shown in Fig. 5 and Table 1. Model 3 was faster than Model 2 while it also achieved a higher performance compared to Model 2.

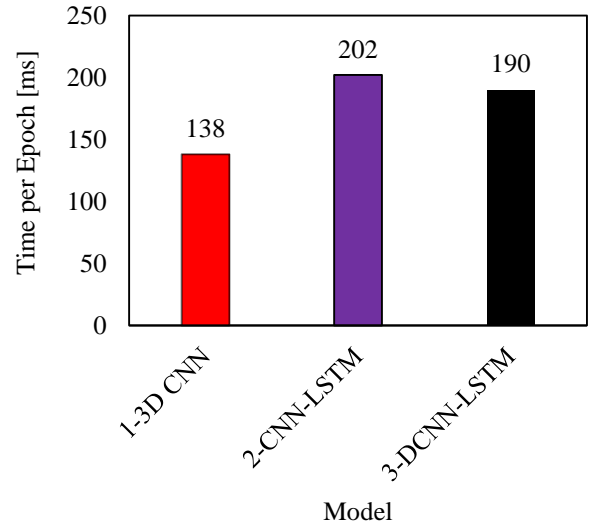

**Fig. 6** Time per epoch for all models including Model 1, 3D Convolutional Neural Network (3D CNN); Model 2, Convolutional Neural Network-Long Short-Term Memory (CNN-LSTM); and Model 3, Differential Convolutional Neural Network-Long Short-Term Memory (DCNN-LSTM)

## 4. DISCUSSION

Integrity of the distal tibiofibular joint is important for conferring stability to the ankle mortise, and thereby plays a crucial role in weight transmission during physical activity. Detection of syndesmotic instability when it exists, and subsequent restoration of mortise integrity, are therefore essential for preventing future degenerative changes of the ankle joint[16,17]. The objective of this research was to train and compare three DL models to detect syndesmotic instability using 3D WBCT images of patients with unilateral syndesmotic instability. Our DCNN-LSTM model was found to exhibit the highest performance, rendering it a promising method to equip clinicians with a better and faster diagnostic tool.

While direct visualization through open or arthroscopic assessment has been purported to be the most accurate diagnostic method to evaluate syndesmotic instability, identifying an equally accurate non-invasive method has remained somewhat elusive to orthopedic care providers [18]. WBCT, as a new imaging modality that provides a dynamic 3D ankle view under physiologic load, is being used to assess different syndesmotic measures including anteroposterior and lateral translation[19], fibular rotation[20], 2D area assessment[21], and 3D volume measurements[3,5]. In a study by Ashkani-Esfahani et al., the predictive values of 2D area and 3D volume measurements were compared[5]. They reported the highest sensitivity (95.8%), and specificity (83.3%) for the 3D volumetric measurement up to 5 cm above the tibia plafond with an interobserver correlation coefficient of 94%. Although 3D volume measurement seems to be a promising and rapidly evolving method for more accurate noninvasive detection of even subtle syndesmotic instability, the time-consuming nature of this method in conjunction with the relatively poor familiarity of clinicians regarding use of 3D measurement tools—and thus the higher propensity for measurement bias—has prevented most care providers from incorporating this method [5].

Clinicians benefit from test simplicity when it can occur without sacrifice of speed, economy, or precision. Use of 3D WBCT volumetric analysis for the syndesmosis remains no exception to this rule. In this study, we developed three DL models for analyzing WBCT scans to recognize syndesmosis instability. These three models included two state-of-the-art models (Models 1 and 2) and a new model (Model 3) that we introduced in this study. We found that Model 1 could not learn any feature for recognizing syndesmosis instability. Model 3 achieved higher accuracy than Model 2 while being faster. Model 3 was able to correctly diagnose all the patients with syndesmotic instability in the test subset and only misclassified one patient. Model 3 also achieved high sensitivity (100%) and specificity (90%) than what is reported in the literature for the 3D measurement method (sensitivity 95.8%, and specificity 83.3%)[5]. Furthermore, Model 3 analyzed a given WBCT scan automatically in a fraction of a second. Although these results show that DL can successfully detect even subtle increased syndesmotic volume that can potentially result in syndesmotic instability, the fact that DL models should be trained on large databases to be generalizable cannot be overlooked. Our study should therefore be considered a pilot for future study of this condition using a more robust database.

      A clear limitation of this study, therefore, remains the relatively small size of its dataset. Since WBCT is a relatively new device and as such the number of studied patients are as yet not high, we are not yet able to provide greater database analytics for the purposes of this investigation. However, our future goal is to expand our data and retrain our model that exhibited the highest performance, following which we intend to perform internal and external validation. Another limitation of this study was the fact that different weight initialization methods were used for Model 1 as compared to Models 2 and 3. Weight initialization sets the weights of a neural network to random values as the starting point for model optimization. An alternative approach to random weight initialization is transfer learning, which involves reusing the weights of a well-trained network to initialize a new network's weights[22]. In this study, Model 2 and Model 3 were created using a transfer learning approach that implemented CNN models trained on the ImageNet dataset as the base model. ImageNet dataset contains over 15 million labeled non-medical natural images[23]. In contrast, Model 1 was trained using random weights with Gaussian distribution. We could not use the same transfer learning approach for Model 1,

because it had a 3D convolution base layer as opposed to models 2 and 3 that were trained on the ImageNet dataset with a 2D convolution layer. Differences in the weight initialization methods may have contributed to the relatively poor performance of Model 1, because it was only trained by the dataset available in this study, as opposed to Model 2 and Model 3 that used a well-trained base CNN (VGG-16) as a feature extractor.

## 5. Conclusion

In this study, we developed a highly accurate and expedient DL model for detection of the unstable syndesmosis using 3D WBCT scans. This method achieved higher performance than the state-of-the-art 3D measurement method while being faster and eliminating the risk of measurement bias. Future steps will include gathering a larger patient dataset in order to increase the validity and reliability of our model. Using heat maps to highlight the level and location of the abnormality can be beneficial in building trust and confidence. Internal validation and external validation of our model using datasets from other centers should help further validate this method and may convince other care providers to use it as a potentially revolutionary new diagnostic tool.

**Acknowledgment**

No external funding was used for this study. The authors have no relevant conflict of interest.